\documentclass[lettersize,journal]{IEEEtran}
\usepackage{amsmath,amsfonts}
\usepackage{algorithmic}
\usepackage{algorithm}
\usepackage{array}
\usepackage[caption=false,font=normalsize,labelfont=sf,textfont=sf]{subfig}
\usepackage{textcomp}
\usepackage{stfloats}
\usepackage{url}
\usepackage{verbatim}
\usepackage{graphicx}
\usepackage{cite}
\usepackage{flushend}
\usepackage{comment}

\usepackage{xcolor} 

\begin{document}

\title{BSoNet: Deep Learning Solution for Optimizing Image Quality of Portable Backscatter Imaging Systems}

\author{Linxuan Li, Wenjia Wei, Yunfei Lu, Wenwen Zhang, Yanlong Zhang, Wei Zhao

\thanks{Linxuan Li, Wenjia Wei and  Wenwen Zhang are with the School of Physics, Beihang University, Beijing, China. (e-mail: zy2219105@buaa.edu.cn, weiwenjia@buaa.edu.cn, and wwzhang@buaa.edu.cn).}
\thanks{Yunfei Lu and Yanlong Zhang are with the Beijing Love Wisdom Fashion Technology Co., Ltd., Beijing, China. (e-mail: ucasluyunfei@163.com, zhangylong@163.com).}
\thanks{Wei Zhao is with the Tianmushan Laboratory and Hangzhou International Innovation Institute, Beihang University, Hangzhou, China, and also with the School of Physics, Beihang University, Beijing, China. (e-mail: zhaow20@buaa.edu.cn).}

\thanks{The corresponding author is Wei Zhao.}}

\markboth{Journal of \LaTeX\ Class Files,~Vol.~14, No.~8, August~2021}%
{Shell \MakeLowercase{\textit{et al.}}: A Sample Article Using IEEEtran.cls for IEEE Journals}

\maketitle

\begin{abstract}
Portable backscatter imaging systems (PBI) integrate an X-ray source and detector in a single unit, utilizing Compton scattering photons to rapidly acquire superficial or shallow structural information of an inspected object through single-sided imaging. The application of this technology overcomes the limitations of traditional transmission X-ray detection, offering greater flexibility and portability, making it the preferred tool for the rapid and accurate identification of potential threats in scenarios such as borders, ports, and industrial nondestructive security inspections. However, the image quality is significantly compromised due to the limited number of Compton-backscattered photons. The insufficient photon counts result primarily from photon absorption in materials, the pencil-beam scanning design, and short signal sampling times. It therefore yields severe image noise and an extremely low signal-to-noise ratio, greatly reducing the accuracy and reliability of PBI systems. To address these challenges, this paper introduces BSoNet, a novel deep learning-based approach specifically designed to optimize the image quality of PBI systems. The approach significantly enhances image clarity, recognition, and contrast while meeting practical application requirements. It transforms PBI systems into more effective and reliable inspection tools, contributing significantly to strengthening security protection.
\end{abstract}


\section{Introduction}
\IEEEPARstart{I}{n} the era of globalization, the tight integration between nations in political, economic, and other fields has brought unprecedented connectivity. However, this interconnectedness has also intensified frictions and conflicts, particularly in areas such as culture and religion, leading to frequent terrorist attacks. As a result, public safety has become a major societal concern~\cite{cirdei2019impact}. Rapid and accurate detection of potential threats and contraband is especially crucial in critical areas like borders, ports, and important public places.

Traditional security screening technologies, while crucial for security checks, 
have shown limitations in addressing complex and evolving security threats. In particular, transmission X-ray imaging, although effective for detecting metal objects such as firearms and knives, has limited capabilities for non-metallic materials like explosive devices and plastic weapons, which absorb X-rays less. This results in insufficient visual contrast for such items in transmission X-ray images, making them difficult to detect~\cite{Mei2015}. Another significant limitation is the requirement for a strict operational environment, where the X-ray source and detector must be positioned on opposite sides of the object being examined. This setup brings considerable challenges in some practical applications, such as industrial security, where inspecting large structures, complex terrains, or objects that are difficult to access from both sides. Inspection technology is needed to meet the demands of complex security checks.
For example, inspecting underground pipelines or assessing road quality using transmission imaging would necessitate excavation to place the equipment underground, significantly increasing inspection complexity and cost. In specialized fields such as the military, public security, and customs, the application of transmission imaging and other conventional inspection technologies is also limited when dealing with large, thin-walled, multi-layered, or geometrically complex objects. Therefore, there is a need for a more flexible and efficient inspection technology to meet the demands of complex security checks.

To overcome the limitations of traditional security inspection technologies, Compton backscatter imaging~\cite{Yan2014} (backscatter imaging) has increasingly gained attention as an emerging non-destructive detection method. Its core principle lies in utilizing scattered photons with angles greater than 90 degrees to obtain material information. Under the same X-ray energy conditions, high atomic number materials primarily interact with rays through the photoelectric effect, while for low atomic number materials, the Compton effect is more significant. Therefore, backscatter imaging technology is particularly suited for distinguishing materials with low atomic numbers, making it an ideal choice for detecting contraband like drugs and explosives.

Despite Compton backscatter  imaging technology demonstrating its unique advantages, it also has certain limitations. Backscatter imaging relies on the scattered photons produced when an object is irradiated with X-rays. However, due to the inherently weak backscatter signal, the number of scattered photons decreases as the detection depth increases, and the likelihood of multiple scattering increases. This leads to a limited number of scattered photons captured by the detector, resulting in a high level of noise in the final detection. Therefore, to enhance the backscatter signal and improve the practical application and reliability of portable backscatter imaging systems (PBI), it is necessary to employ signal enhancement techniques.

Traditional image enhancement methods can improve the quality of backscatter images to some extent, but these methods often rely on manual parameter tuning and lack adaptability. This is particularly challenging in the application of PBI, where the inherent weakness of the backscatter signal and high noise interference make it difficult for traditional methods to effectively address complex noise patterns. In recent years, deep learning-based techniques have gradually become the mainstream approach to improving image quality~\cite{lecun2015deep}. Deep learning models can adaptively learn and optimize complex noise patterns, thus overcoming the limitations of traditional methods. On this basis, self-supervised learning, a strategy that generates supervision signals from the input data itself without relying on high-quality labeled data, has gained increasing attention~\cite{ericsson2022self}. Given the difficulty in obtaining high-quality labeled data in backscatter imaging, self-supervised learning provides significant support for deep learning applications in PBI.

This work introduces a network (BSformer) specifically designed for backscattered image restoration, and a Resolution Adaptive Network (RANet). The BSformer and RANet are incorporated to constitute the overall framework referred to as BSoNet.   
BSoNet can better adapt to the diverse and complex real-world application scenarios of PBI. To the best of our knowledge, this study marks the first application of deep learning technology to the task of image quality optimization for PBI, pioneering a new research direction in this field. The main contributions of this paper are as follows:

\textbf{(1). A powerful model BSformer with global and local information fusion capabilities:}  We propose a powerful deep network model, BSformer, which combines the global information processing capabilities of the Transformer with the local feature extraction advantages of CNNs. This fusion not only allows the model to understand image content at a global level and capture dependencies over large areas, but also enables it to finely process local details within the image, thereby achieving more precise and comprehensive backscatter image quality optimization.

\textbf{(2). Flexible adaptive processing to ensure the practicality of the model:} In the BSoNet framework, the adaptive processing mechanism benefits from the proposed RANet. This network is dedicated to ensuring the consistency of image size and structure before and after optimization. It precisely adjusts the input images to the standard size required by the BSformer and finely restores them to their original dimensions after optimization.
As a result, BSoNet can seamlessly adapt to various scanning conditions and environmental changes encountered by PBI in practical use, such as differences in image size and quality caused by scanning speed and scanning duration. Therefore, in scenarios requiring rapid and precise identification of potential threats, BSoNet's flexible adaptive processing capability is able to play a crucial role.

\textbf{(3). Enhanced application of self-supervised learning strategies:} 
This study employs a self-supervised learning strategy, noise2void~\cite{krull2019noise2void} to tackle the challenge of lacking clean training labels for PBI. Based on this strategy, we further impose additional noise augmentation to the original backscatter data during the training process, compelling the model to actively adapt and improve optimizing performance under complex data conditions. This training method significantly enhances the model's adaptability and processing capability in complex noise environments, thereby effectively improving the performance of backscatter image quality optimization under label-free conditions. This provides a new solution to deep learning-based PBI image optimization.

\section{Research Basis}
\subsection{Imaging Principle of Backscatter Systems}
Backscatter imaging is based on the Compton scattering effect. When an incident photon interacts with the electrons of the target material, it transfers part of its energy to the electron, resulting in a change in the photon's direction and energy. This phenomenon is known as Compton scattering.

\begin{figure}[!ht]
\centerline{\includegraphics[width=\columnwidth]{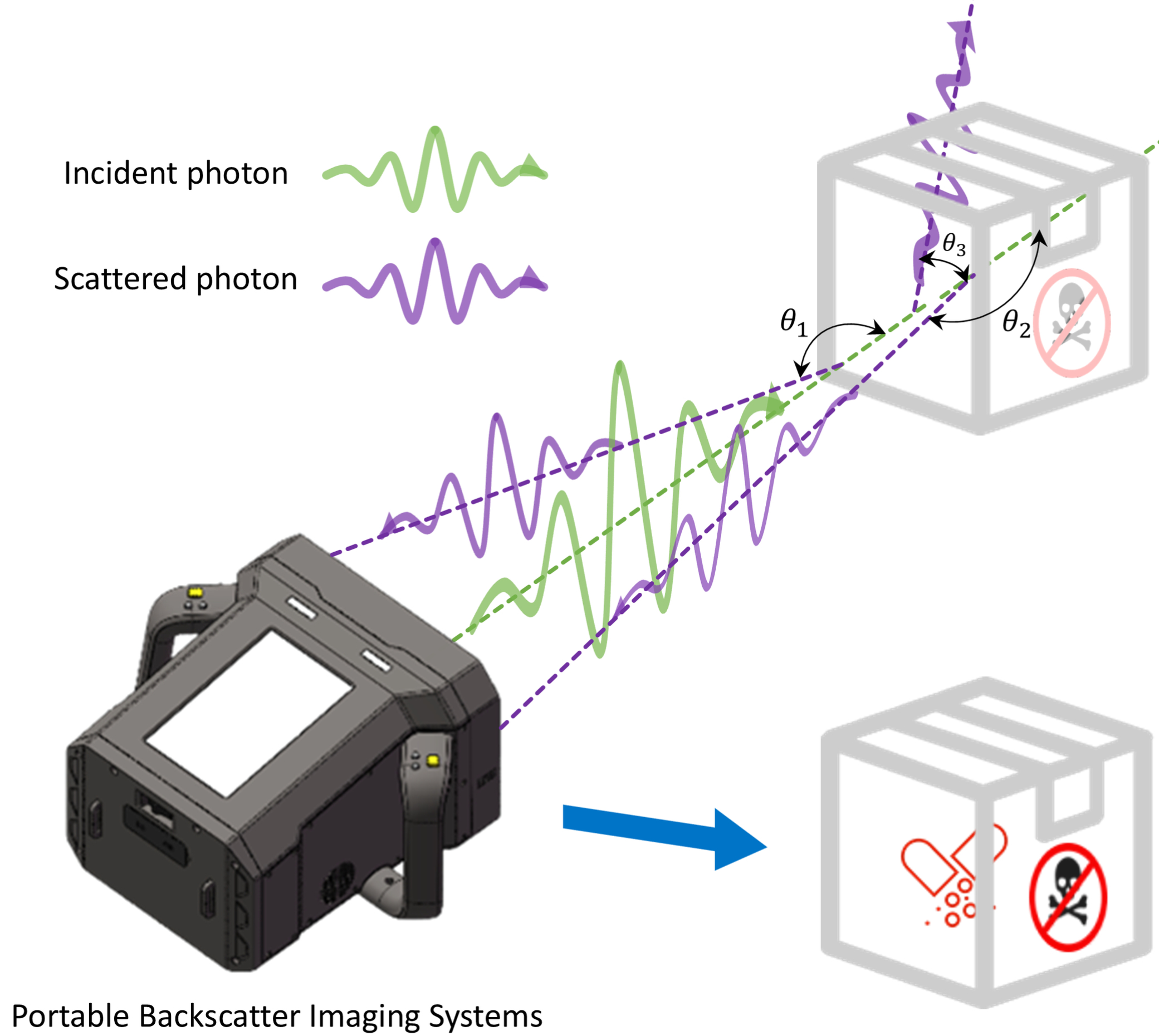}}
\caption{Schematic diagram of the portable backscatter imaging system, demonstrating  the Compton scattering effect caused by the incident photon interacts with the electrons of the target material, resulting in forward scattered photons (at an angle of ${{\theta }_{3}}$) and backscattered photons (at angles of ${{\theta }_{1}}$ and ${{\theta }_{2}}$). The backscattered photons are received by the device detector.}
\label{fig1}
\end{figure}

Depending on the movement direction of the scattered photons, Compton scattering can be classified into forward scattering and backscattering.  When the scattering angle $\theta$ is greater than 90 degrees (as shown by ${{\theta }_{1}}$ and ${{\theta }_{2}}$ in Fig. \ref{fig1}), the scattered photon and the recoil electron move in opposite directions, which is known as Compton backscattering. This is the core signal source of backscatter imaging.

Backscatter imaging systems can achieve position resolution in two ways: one is by adding collimators in front of each detector pixel, so that each pixel detector only receives a certain position on the object pointed by the corresponding collimator. However, this approach leads to most X-rays being absorbed by the collimator. Furthermore, due to the inherently weak scatter signal, the intensity of the received scattered X-rays is very low. The second method is the flying spot scanning technology, only one pencil beam of rays is irradiated onto the object at a certain moment during the scanning process.  The position of the beam is determined by a synchronization system. This method avoids the limitations of collimators, increases photon reception efficiency, and significantly reduces the emitted X-ray dose, ensuring the safety of the operator.

The portable backscatter imaging system uses a specific collimation mechanism to form a pencil beam (also called a flying spot) and project it onto the surface of the object being inspected. By precisely controlling the movement position of the X-ray beam and combining it with the backscattered X-ray signals received by the detector, the flying spot movement path is scanned point by point. At the same time, the entire device moves synchronously in the direction perpendicular to the flying spot scanning plane, ultimately achieving a point-to-line-to-surface imaging process. Ultimately, the imaging system displays the two-dimensional density distribution information of the shallow layer of the inspected object.

\subsection{Research Conditions}
\begin{figure}[!ht]
\centerline{\includegraphics[width=\columnwidth]{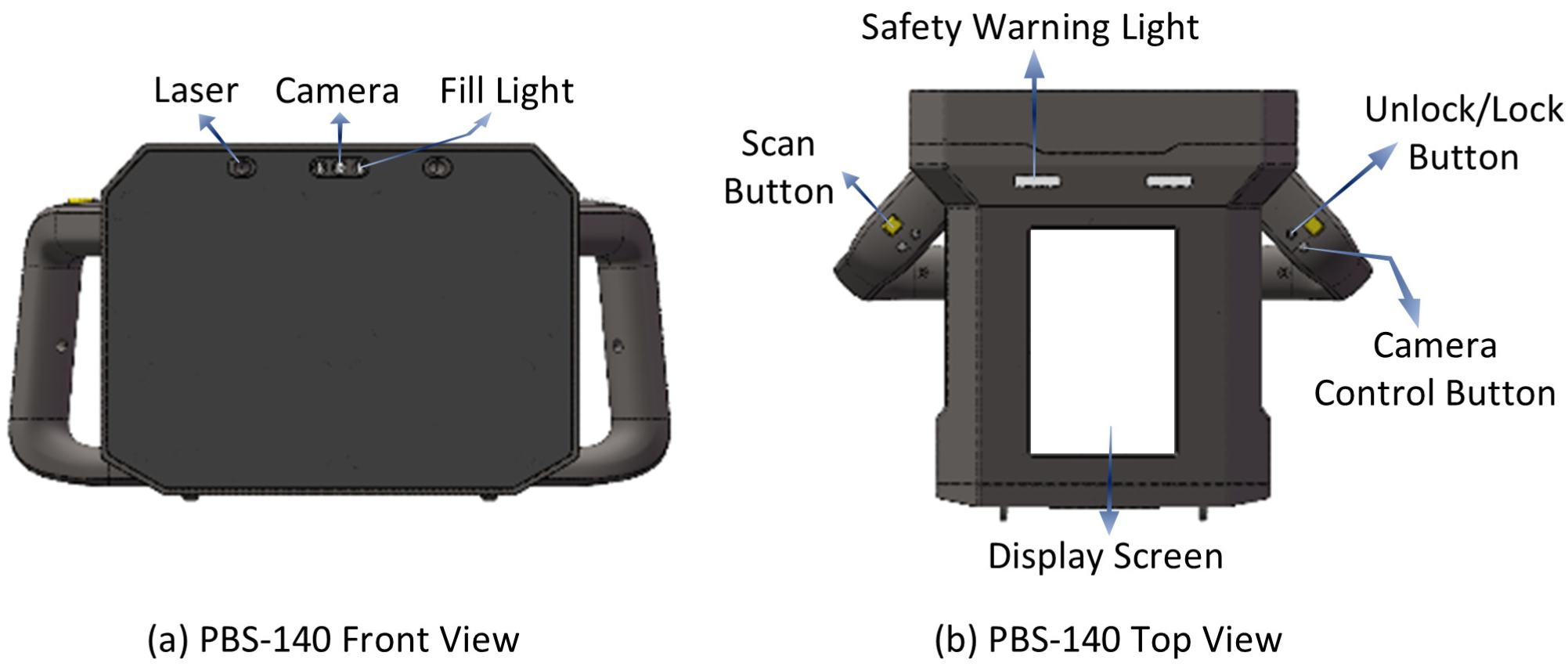}}
\caption{Schematic diagram of the  structure and some components of the PBS-140 portable backscatter imaging system.  (a) Front view  of PBS-140. (b) Top view  of PBS-140.}
\label{fig2}
\end{figure}

This study employs the PBS-140 portable backscatter imaging system, independently developed by Beijing Love Wisdom Fashion Technology Co., Ltd. (as is shown in Fig. \ref{fig2}). The device has successfully passed the rigorous certification of the China Ministry of Public Security's Safety and Police Electronic Product Quality Testing Center, 
making it highly effective in detecting concealed items. The device holds independent intellectual property rights and multiple invention patents, with broad application prospects in public security, border control, and other fields. The PBS-140 is designed with a focus on portability and efficiency, utilizing flying spot scanning technology to effectively enhance photon reception efficiency while reducing radiation dose.

The device supports a maximum tube voltage of 140kV and a maximum tube current of 50$\mu\text{A}$, allowing users to flexibly select the appropriate settings based on the actual detection environment. Specifically, a higher tube voltage can enhance the penetrating power of X rays, making it suitable for detecting high-density or thick objects; however, it may also reduce the differences in absorption among various substances and thus lower image contrast. In contrast, a lower tube voltage helps improve image contrast and is more suitable for detecting low-density substances, though its penetrating power is comparatively weaker. A higher tube current increases the number of X ray photons, thereby enhancing the signal to noise ratio and detail clarity of the image, but results in a higher radiation dose; conversely, a lower tube current helps control the radiation dose and reduce radiation risk, but may affect the image signal-to-noise ratio and detail clarity to some extent.

In addition, the key publicly available technical specifications of the PBS-140 include its scanning capability (with a maximum single scan duration of 30 seconds and a scanning speed of no less than 20 cm/s), steel plate penetration (not less than 5 mm), and imaging resolution (backscattered ray resolution is 1 mm, backscattered ray detection capability is 0.6 mm). The device also offers a storage capacity of at least 20,000 images. The PBS-140 is equipped with sodium iodide scintillation detectors combined with a photomultiplier tube system, which achieves high sensitivity and strong signal amplification capabilities.
In terms of radiation safety, the device performs  well: the dose equivalent rate measured at approximately 2 cm from the rear of the device and around the handle area is below  0.25 $\mu\text{Sv/h}$, and under maximum radiation conditions, the radiation dose per scan remains below 5 $\mu\text{Sv/h}$, thereby fully ensuring the safety of operators and the public.

In each experiment, detailed information regarding the tube voltage and tube current was provided, aiming to offer as comprehensive a technical reference as possible for researchers and to support the progress of related research.

\section{Related Works}
In the application of PBI, deficiencies in factors such as the stability of the radiation source, the uniformity of the detector response, and the mechanical precision of the equipment often lead to inconsistencies in image quality during the imaging process. These inconsistencies can manifest as streak artifacts, increased noise levels, and decreased signal-to-noise ratio, affecting the overall image quality. Furthermore, due to the inherently weak backscatter signal, especially when imaging the deeper structures of objects, these issues become more pronounced. Therefore, image enhancement techniques are crucial for improving the performance of PBI. The noise in backscatter images can be categorized primarily as impulse noise and Gaussian noise~\cite{Wang2013}. There are many ways to deal with these two types of noise.

Noise in essence can be either additive or multiplicative. 
However, in most cases, the noise is considered to be independently and identically distributed, with a mean of zero~\cite{goyal2020image}.
The imaging model can be approximately represented by \eqref{eq1}:
\begin{equation}
\label{eq1}
f(x)=u(x)+n(x),x\subset X,X\subset {{Z}^{2}},
\end{equation}
where $x$ represents the coordinate position of a specific pixel in the image, $X$ represents the set of coordinates for all pixels in the image, $Z$  represents the set of all integers, ${Z}^{2}$  represents the set of all possible integer coordinates,
$u(x)$ denotes the pixel value of the original noise-free image at position $x$, $n(x)$ represents the noise at position $x$, and $f(x)$  is the pixel value of the image containing noise at location $x$, representing the real measurement. Under this formulation, if the noise $n(x)$ can be accurately determined, the original image is able to be recovered by subtracting the noise $n(x)$ from the input image $f(x)$. However, in practice, unless the noise generation mechanism is explicitly known, it is difficult to isolate the noise component from the image.

Over the years, traditional denoising methods, machine learning methods,  deep learning approaches, and other mathematical methods have been applied to the task of image denoising. Among these, traditional denoising methods and deep learning methods have been most commonly used, yielding successful results~\cite{yapici2021review}. In this section, we provide a comprehensive review of traditional image denoising methods and the recently emerged deep learning-based methods. Additionally, we will outline the methods proposed in this study.

Traditional denoising methods are classified into four categories based on their characteristics~\cite{goyal2020image}:

(1) Spatial domain filtering. Spatial domain filtering can be further divided into two subcategories: local filtering and non-local filtering. Common local filters, such as the mean filter~\cite{weizheng2020digital}, gaussian filter~\cite{hsiao2007generic}, bilateral filter~\cite{tomasi1998bilateral}, median filter~\cite{lee1985generalized}, and SUSAN filter~\cite{smith1997susan}, primarily rely on the statistical properties of  neighborhood pixels. They achieve image smoothing by calculating the average or weighted average of neighboring pixels. Non-local filters, such as non-local means (NLM)~\cite{buades2005review}, utilize the correlation between the overall pixels across the image for denoising. The core idea of NLM is to find pixels or regions within the entire image that are similar to the target pixel and use the weighted average of these similar regions to reconstruct the target pixel value. Additionally, researchers have developed various improved algorithms based on the NLM principle, aiming to further enhance denoising performance and optimize computational efficiency, including INLM~\cite{goossens2008improved}, NLMPG~\cite{xu2017remote}, and others. However, despite the simplicity and effectiveness of spatial domain filtering, in strong noise scenarios, the correlation between pixels is severely damaged by high levels of noise, and its denoising performance may be seriously degraded. 

(2) Transform domain filtering. As a transform domain method, Fourier transform~\cite{sifuzzaman2009application} was first introduced as a basic processing method. With advances in technology and a deeper understanding of image characteristics, various transform domain methods have been developed, such as the discrete cosine transform (DCT)~\cite{lebrun2012secrets}, wavelet-based denoising methods~\cite{kingsbury1998dual}, and the advanced block-matching and 3D filtering (BM3D)~\cite{dabov2007image}. Additionally, independent component analysis (ICA)~\cite{stone2004independent} and principal component analysis (PCA)~\cite{shlens2014tutorial} have also been widely applied as effective transform techniques. These transforms reveal the sparse nature of image signals in the transform domain, that is, the main information in an image can be effectively represented by a small number of non-zero coefficients in the transform domain. This sparsity not only simplifies signal representation but also makes the distinction between signal and noise clearer, making it possible for precise denoising. However, transform domain methods also have certain limitations. For example, wavelet transforms are less effective in handling smooth transition regions, and discrete cosine transforms have difficulty accurately capturing sharp edges or singularities in images. Moreover, the performance of these methods largely depends on the accurate setting of parameters.

(3) Other domain methods. 
These methods exploit deep statistical information contained in image data for image denoising. For example, methods based on random fields and dictionary learning have demonstrated unique advantages in handling image noise, and Markov Random Fields (MRF) model the spatial dependencies between pixels to restore the overall characteristics of the image~\cite{rangarajan1995markov}. Sparse representation and dictionary learning methods provide a new perspective for efficient image denoising by mining sparse patterns from image data and constructing optimized representation dictionaries. The core of these methods lies in identifying the smallest set of coefficients that best represent the image content through the optimization process, effectively separating noise and signal to achieve denoising~\cite{elad2006image}. 

(4) Hybrid denoising techniques. These methods combine the strengths of multiple domains to overcome the limitations of single-domain methods. For example, spatial domain filtering can effectively preserve edge information but may result in excessive smoothing of low-contrast details. Transform domain methods, while advantageous in representing textures and low-contrast information, may produce ringing effects near edges. Based on these characteristics, Knaus and Zwicker~\cite{knaus2013dual} proposed an innovative dual-domain image denoising method (DDID) that combines bilateral filtering in the spatial domain with short-time Fourier transform and wavelet shrinkage techniques in the transform domain. This method not only retains low-contrast details but also ensures edge sharpness. However, designing and implementing such hybrid methods requires precise balance to fully leverage the advantages of different techniques without introducing new issues. Additionally, careful selection of various parameters used in the denoising process is essential to avoid excessive smoothing or the formation of artifacts near edges.

In recent years,  
deep learning has found widespread applications across disciplines~\cite{lecun2015deep,shen2019patient}, such as object detection~\cite{redmon2016you,zhao2019markerless}, image restoration~\cite{ulyanov2018deep}, semantic segmentation~\cite{long2015fully}, and image synthesis~\cite{wan2023quantitative}. The field of image denoising has similarly benefited from these advancements, with deep learning surpassing the boundaries of traditional denoising algorithms and extending to the handling of complex noise patterns in images captured under various conditions. This is attributed to the superior ability of deep learning to automatically extract features, overcoming the limitations of traditional algorithms in noise identification and processing.

According to the type of noise, deep learning image denoising algorithms can be divided into three categories~\cite{tian2020deep}:
(1) Denoising for additive white Gaussian noise. This type of denoising method primarily targets white Gaussian noise in images.
Herein the denoising algorithm needs to identify and remove this Gaussian noise while preserving as much image detail and quality as possible. For example, Burger~\textit{et al}. proposed an image denoising method based on a multi-layer perceptron (MLP). By training on a large-scale image dataset, it successfully proved that it could match or even surpass the most advanced image denoising technology at the time, such as BM3D~\cite{burger2012image}. Zhang~\textit{et al}. introduced a feedforward denoising neural network (DnCNN) that combines batch normalization and residual learning techniques for image denoising~\cite{zhang2017beyond}. Additionally, in another study, Zhang~\textit{et al}. trained a highly flexible and effective convolutional neural network (CNN) denoiser, establishing a benchmark deep denoiser prior. This deep denoiser prior was then integrated as a module into a semi-quadratic splitting-based iterative algorithm to solve various image restoration problems. (2) Denoising images with real noise. This type of denoising targets noise real present in images acquired in real-world environments. Real noise can originate from various factors, such as the the photosensitive element of the shooting equipment or ambient lighting conditions, and it is typically complex and diverse in both type and intensity. Therefore, denoising algorithms must adaptively identify and handle various complex noise patterns. For example, Yan~\textit{et al}. extracted noise maps directly from noisy images to achieve unsupervised noise modeling, enabling unpaired real noise image denoising~\cite{yan2019unsupervised}. Syed Waqas Zamir~\textit{et al}. proposed the CycleISP model, which simulates camera imaging processes to generate an arbitrary number of realistic noisy images for training image denoising models. This strategy significantly improves the image recovery performance of the CycleISP model on the real camera benchmark datasets and also significantly reduces the number of parameters of the model~\cite{zamir2020cycleisp}. 
(3) Denoising blind noise images. Blind denoising refers to denoising images without specific knowledge of the noise type and level. The denoising algorithm needs to be highly adaptive and flexible to achieve effective denoising in the absence of prior knowledge of the noise. For example, Yu~\textit{et al}. proposed a deep iterative down-up CNN for image denoising, which can repeatedly reduce and increase the resolution of feature maps. This method can handle various noise levels with a single model, without requiring noise information as input~\cite{yu2019deep}. Zhang~\textit{et al}. proposed a fast and flexible denoising CNN, FFDNet, which effectively processes images with different noise levels by inputting adjustable noise level maps. FFDNet can efficiently remove spatially varying noise and achieves a good balance between inference speed and denoising performance. Even on a CPU, FFDNet outperforms the benchmark BM3D in terms of speed without sacrificing denoising capabilities~\cite{zhang2018ffdnet}.

The recently proposed Transformer paradigm has marked a transformative 
breakthrough 
in natural language processing and advanced vision tasks~\cite{vaswani2017attention}. This architecture utilizes a self-attention mechanism to efficiently extract intrinsic features of data. The main advantage of this mechanism lies in its ability to understand long-range dependencies between elements within a sequence. Compared to traditional neural network architectures, Transformer is more efficient and accurate in handling complex data, thus showing extensive and tremendous potential in various applications of artificial intelligence. Particularly in the field of image restoration, Uformer, a novel architecture based on Transformer, has directly occupied the pinnacle in areas such as image denoising, image deraining, and image deblurring~\cite{wang2022uformer}.

\begin{figure*}[!ht]
\centerline{\includegraphics[width=7in]{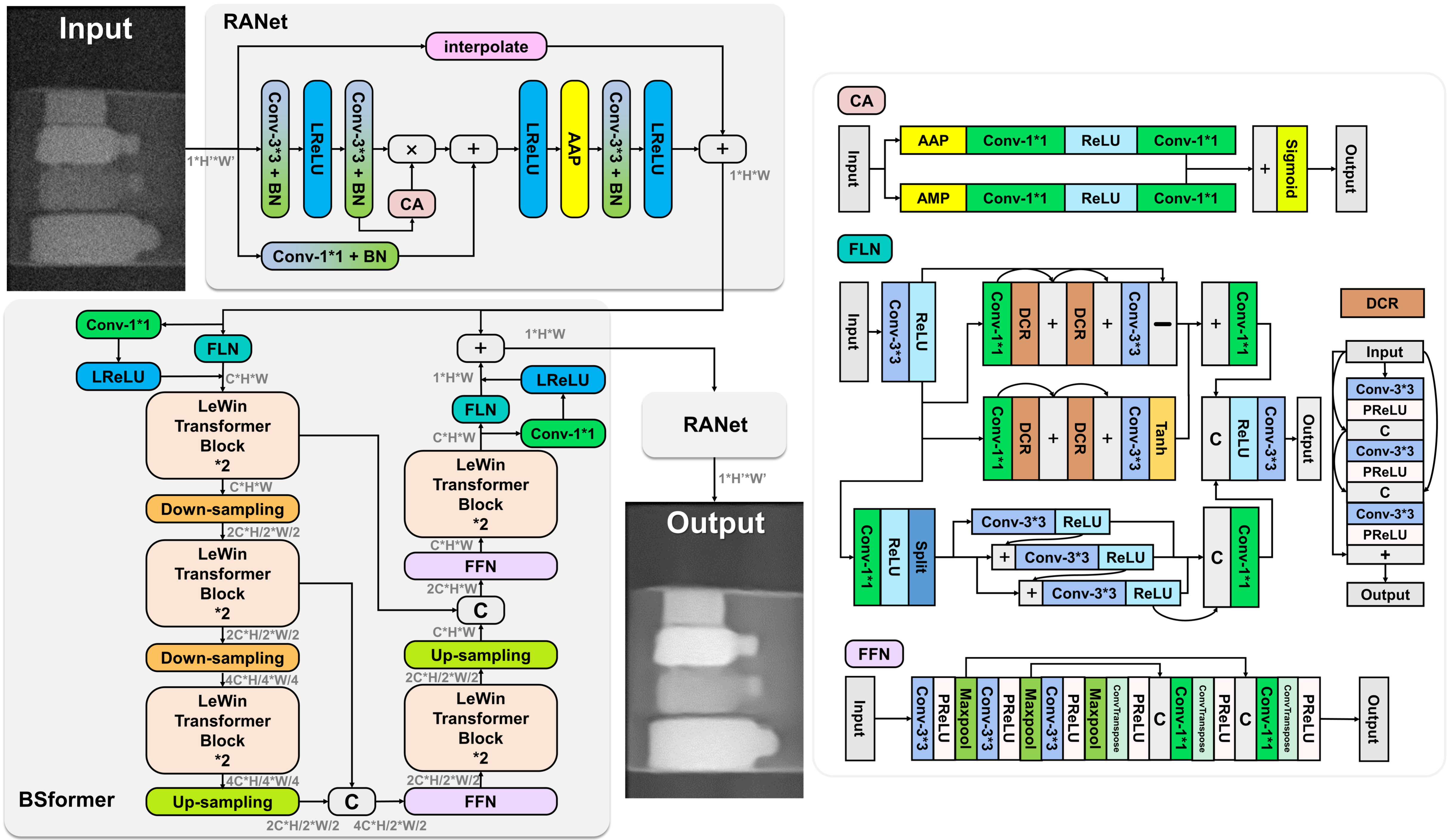}}
\caption{BSoNet, a deep learning-based imaging quality optimization framework for PBI and its key components. RANet: Resolution Adaptive Network. BSformer: Backscatter Optimization Transformer. FLN: Feature Learning Network structure within BSformer. FFN: Feature Fusion Network structure within BSformer. CA: Channel Attention module within RANet. DCR: Dense Connection Block network structure within FLN. AAP: Adaptive Average Pooling layer. AMP: Adaptive Max Pooling layer.}
\label{fig3}
\end{figure*}

\section{Methodology}
In this section, we will provide a detailed description of the overall design process of the PBI image quality optimization network, BSoNet, as well as the components involved.
\subsection{Framework Overview}
In this paper, we propose BSoNet, an innovative deep learning framework designed to improve the quality of backscatter images, particularly improving the performance of PBI across a wide range of application scenarios. This approach fully integrates the global information processing capabilities of the Transformer architecture with the local feature extraction advantages of CNN, offering an efficient solution to common image quality degradation issues in the field of backscatter imaging, such as noise and resolution loss.

Fig. \ref{fig3} demonstrates the overall framework of BSoNet, which adopts a symmetrical resolution adjustment strategy, consisting of three main stages: initial resolution adjustment stage, core optimization stage, and final resolution restoration stage. The three stages work closely together to ensure the high-quality optimization of backscatter images.

In both the initial and final stages of this framework,  the self-developed RANet is deployed to meet the specific needs of different stages. The RANet in the initial stage adjusts the input image to the appropriate size required by BSformer while preserving the original structure and noise distribution of the image as much as possible, thus avoiding additional distortions that might be introduced by resolution adjustments and laying a foundation for deep image optimization.   The RANet in the final stage meticulously restores the BSFormer's optimized image  to the original input dimensions, ensuring consistency between the output and input sizes and further maintaining the integrity of the image structure. The core optimization part is borne by the BSformer, an innovative model that fully integrates the global information processing capabilities of the Transformer architecture with the local feature extraction advantages of CNN, focusing on deep optimization of backscatter images. Through efficient denoising techniques and detail enhancement means, it effectively improves the overall quality of the backscatter images.

BSoNet, through the precise resolution adjustments before and after by RANet and the efficient image optimization capabilities of BSformer, ensures excellent backscatter image optimization performance under various usage conditions. This greatly enhances the applicability of portable backscatter imaging systems in diverse application environments. Furthermore, by integrating enhanced self-supervised learning strategies, BSoNet can effectively learn even in challenging environments lacking clean training labels, exploring new research directions for the application of deep learning in backscatter imaging technology. 

\subsection{Resolution Adaptive Network (RANet)}
In the practical application of optimization methods for PBI, maintaining the original structure of input images under varying conditions is key to ensuring the final image quality. In response to this demand, we developed  RANet, which is committed to preserving the consistency of image dimensions and structure before and after optimization. This is crucial for preserving the authenticity of the imaging results.  Its overall structure is shown in the corresponding part of Fig. \ref{fig3}. 

The architecture of RANet is meticulously designed, integrating end-to-end bicubic interpolation residual connections and feature learning branches. The interpolation branch provides a stable learning foundation for the feature learning branch, effectively preserving the original structure of the image during size changes and preventing potential information loss caused by resolution adjustments. The Channel Attention (CA) mechanism further enhances the model's ability to capture important features. The combined use of adaptive average pooling layers and interpolation residual connections allows RANet to adaptively adjust image dimensions without losing image quality or altering the original noise patterns.

Through this structural design, RANet ensures that images under different input conditions can be accurately adjusted to the unified standard dimensions required for deep optimization by BSformer and seamlessly revert to their original input dimensions after processing. This feature is crucial for the practical application of PBI. Therefore, RANet provides a robust foundation for our BSoNet framework, enabling it to handle diverse inputs while ensuring the authenticity and reliability of the output images.

\subsection{Backscatter Optimizing Transformer (BSformer)}
Although deep learning has achieved remarkable success in various computer vision tasks,  its application in backscatter image quality optimization has not been reported. Our novel network, BSformer, is a first step in this direction. Inspired by the Uformer model, BSformer integrates the global information processing capabilities of the Transformer architecture with the local feature extraction advantages of CNN, aiming to comprehensively enhance the quality of backscatter images.  Its structure is shown in the corresponding part of  Fig. \ref{fig3}. 

At the front and end of the BSformer model, we designed the Feature Learning Network (FLN). Its structure is shown in the corresponding part of  Fig. \ref{fig3}. FLN consists of two branches. The design inspiration of the upper branch comes from the two sub-branches of the underlying structure of the dense self-guided wavelet network~\cite{liu2020densely}, it has been proven that the residual branch focuses on processing bright areas in the image, while the end-to-end branch is more suited for addressing noise in dark areas. The lower branch adopts a multi-scale feature extraction residual structure, by dividing the feature maps into multiple subsets and processing them through different convolutional layers, then sequentially adding these processed feature subsets. This approach enhances BSformer's multi-scale feature extraction capabilities while maintaining a low computational burden. The overall structure of FLN consists of well-designed upper and lower branches, each playing a key role.

Additionally, within the BSformer structure,  we designed the Feature Fusion Network (FFN) integrated into the feature processing flow. By integrating multi-scale feature extraction, cross-scale feature fusion, and channel compression methods, the FFN can effectively fuse different levels of image features extracted by BSformer, reduce information loss, and optimize the model's computational efficiency. 

At the input of the model, the FLN first extracts the low-level features of the input image. These features provide rich information for the subsequent LeWin Transformer blocks~\cite{wang2022uformer}, enhancing the model's ability to represent image content. By deeply analyzing the basic structure of the image,  the FLN accurately captures fine textures and key details, which are indispensable for predicting high-quality images.

During the downsampling process, BSformer extracts key macro features by gradually reducing the size of the image features. These macro features are the basis for understanding the image structure and are subsequently passed layer by layer to the LeWin Transformer blocks. Through its multi-head self-attention mechanism of non-overlapping windows, enhances the model's ability to capture global dependencies while preserving critical global details, ensuring information integrity during feature compression. The upsampling stage is a complex feature optimization process. By reapplying the LeWin Transformer blocks at each step of upsampling,  the model can precisely refine and enhance local image details while restoring the image size, ensuring that key visual information is preserved.  Additionally, FFN plays a crucial role in this stage. By  integrating the feature outputs of LeWin Transformer blocks at different levels, the FFN ensures maximum retention and enhancement of details and quality during the spatial dimension restoration process.

Finally, at the end of the model,  FLN undertakes the reconstruction task.  It integrates the global and local features obtained from the downsampling and upsampling processes to finely reconstruct the high-level features and structures of the image. This process helps to restore details that may have been lost during deep processing, and also significantly enhances the contrast and clarity of the image, ensuring the high quality of the final output.

\subsection{Enhanced Application of Self-Supervised Learning Strategies}
In the fields of computer vision and natural language processing, supervised deep learning methods utilizing large-scale labeled  datasets have achieved significant success, especially the application of CNN, which has promoted the rapid development of the visual correlation subfield. However, in certain specialized research or industrial applications, such as backscatter imaging, high-quality labeled data is often difficult to obtain, which limits the application of traditional supervised learning methods.

To overcome this challenge, our framework adopts the Noise2Void self-supervised learning strategy, leveraging the intrinsic statistical properties of the images themselves to guide the learning process.  Using this approach, the model can recover clear image details from noisy images without any clean labels. This strategy significantly expands the applicability of deep learning in data-limited environments, providing a new solution for enhancing image quality in the field of backscatter imaging.

In applying the Noise2Void strategy to backscatter image processing, we first select a certain proportion of target pixels in the image. Then, within the neighborhood of each target pixel, we randomly choose the value of a nearby pixel to replace the target pixel. This method generates a new \textcolor{blue}{``}noisy" image, which is used as the input for model training, while the original image serves as the training target label. To further improve the model's denoising capability, we additionally add standard Gaussian noise of a certain intensity to the training data pairs. This approach not only increases the complexity of the data but also enhances the model's sensitivity to common noise patterns in backscatter images, thereby enhancing  the denoising effect.

\section{Experiments and Results}
\subsection{Data Acquisition and Model Implementation Details}
The data used in this study are all from the PBS-140 portable backscatter imaging system independently developed by Beijing Love Wisdom Fashion Technology Co., Ltd. Our data simulate multiple usage scenarios of the portable backscatter imager, such as detecting potential contraband inside objects, industrial safety inspections, and diverse imaging conducted at different voltages, scanning speeds, and scanning durations. The diverse settings simulate various scenarios that might be encountered in practical operations,  allowing the model to comprehensively learn imaging performance under different technical parameters.  Additionally, the dataset includes test images specifically targeting particular detection indicators, such as spatial resolution tests and penetration tests, which help in model calibration and improving image quality.  A total of 906 raw data images were collected for this study, with 760 used to construct the training dataset and 146 used as the test dataset.

Our model is implemented in PyTorch and trained on a workstation equipped with an NVIDIA RTX 3090 Ti graphics processing unit (GPU) with 24GB of memory. We enhance the training sample by randomly flipping.  
In constructing training data pairs using Noise2Void, we randomly selected 10\% of the target pixels in each image and replace them with values from their neighboring pixels. Additionally, the intensity of standard Gaussian noise added to the training data pairs constructed by Noise2Void is 400.
We train our framework using the AdamW optimizer~\cite{loshchilov2017decoupled}, with the momentum term set to (0.9, 0.999) and the weight decay set to 0.02. We train for a total of 250 epochs and select the model that achieves the best loss. We use a cosine annealing strategy to reduce the learning rate to $1 \times 10^{-6}$, with an initial learning rate of $1 \times 10^{-4}$. In all LeWin Transformer blocks, we set the window size to $8 \times 8$. 

Given that backscatter images are usually accompanied by high noise and weak signals, it is difficult to preserve details and edge information, which is crucial for target recognition. The Charbonnier loss has the following advantages in theory, which can enhance these key image features~\cite{lai2018fast}. The Charbonnier loss is defined as~\eqref{eq3}:
\begin{equation}
\label{eq3}
\ell (\mathbf{I},\widehat{\mathbf{I}})=\sqrt{\|\mathbf{I}-\widehat{\mathbf{I}}{{\|}^{2}}+{{\epsilon }^{2}}},
\end{equation}
where $\mathbf{I}$ represents the target label constructed by Noise2Void, $\widehat{\mathbf{I}}$ is the output predicted by the model, and $\epsilon$ is a small constant (set to $1 \times 10^{-3}$) to ensure differentiability when the error is zero. Compared with the L1 loss, the Charbonnier loss is twice differentiable when the error is close to zero, which ensures smooth gradients and a stable training process. When the error is small, its gradient approximates that of the L2 loss, which can provide higher gradients that facilitate the fine restoration of image details. In regions with larger errors, its gradient approaches that of the L1 loss, reducing outlier penalties and edge blur.

\subsection{Performance Evaluation}
In the practical application of  PBI, the visual optimization effect of the algorithm on raw data is the most important criterion for evaluating its performance. To intuitively demonstrate that our BSoNet can predict images with  competitive quality, we conducted comparative experiments using multiple algorithms,  including ResNet~\cite{he2016deep}, UNet~\cite{ronneberger2015u},  Uformer, as well as traditional methods such as Gaussian filtering, Bilateral filtering, and NLM. For the deep learning methods (ResNet, UNet, Uformer), we adopted hyperparameter settings that are as consistent as possible with those of BSoNet: we uniformly employed the AdamW optimizer (momentum term was set to (0.9, 0.999), weight decay was set to 0.02), utilized Charbonnier Loss as the loss function, set the number of training epochs to 250, and ultimately selected the model based on the best achieved loss. For the traditional methods, Gaussian filtering was implemented using a $7 \times 7$ kernel and a standard deviation of 1.5; for Bilateral filtering, the color-space standard deviation was set to 600, the coordinate-space standard deviation to 8, and the neighborhood diameter was set to 25; for NLM, the filtering strength for each image was determined using the $estimate\_sigma$ function from the scikit-Image library, with the patch size for similarity comparison set to $7 \times 7$ and the search region radius set to 8 pixels.

\begin{figure}[!ht]
\centerline{\includegraphics[width=\columnwidth]{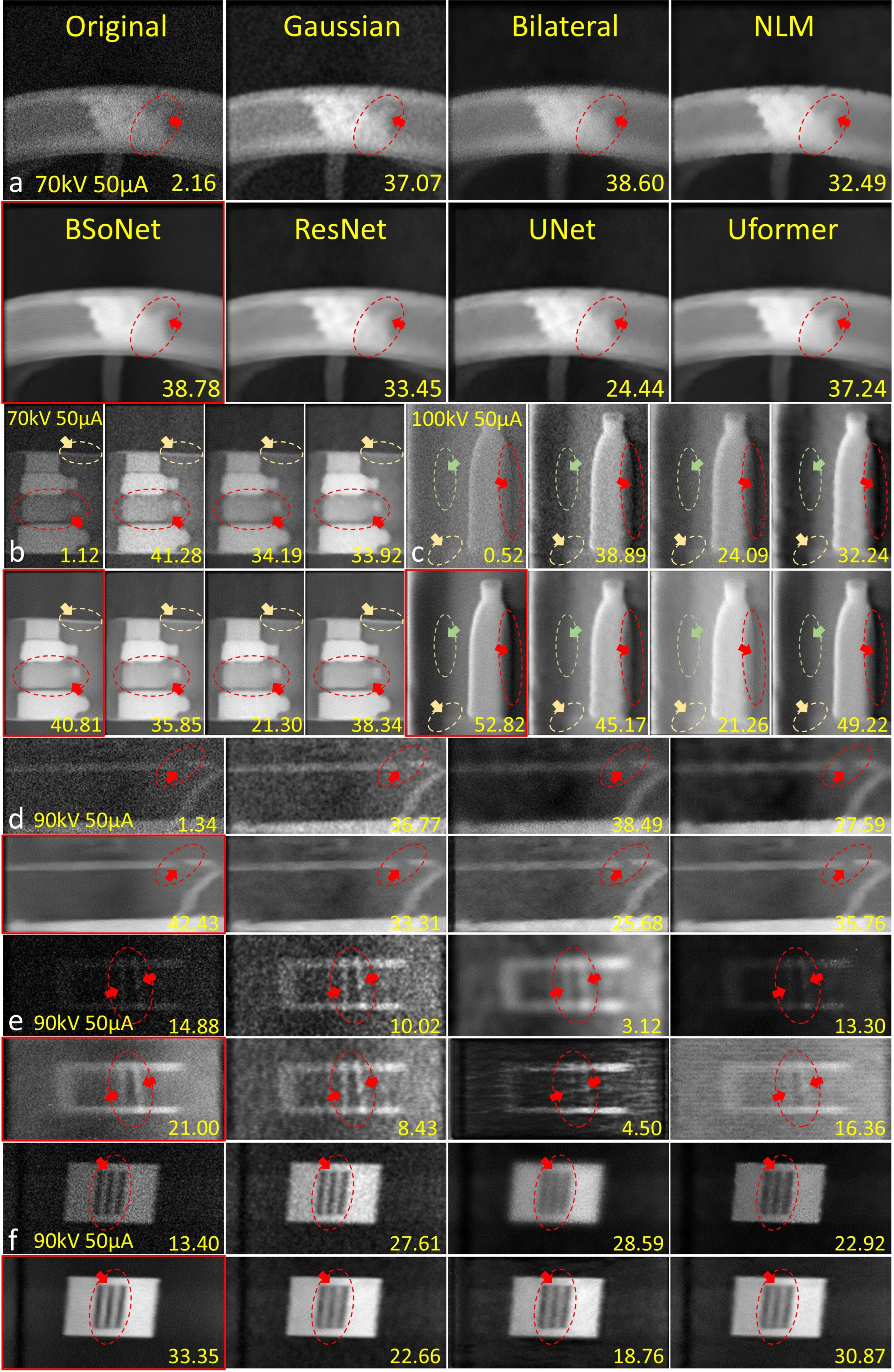}}
\caption{Visual optimization effects of BSoNet and other optimization methods on several representative cases (including quantitative results of local contrast). Each set of images illustrates a specific case. Our method is highlighted with a red box, and the relative position of each method is marked in detail in group (a) to facilitate observation and comparison of the effects of different processing techniques.
}
\label{fig4}
\end{figure}

We selected the following representative cases to showcase the performance of BSoNet:

\textbf{(1). Identifiable Optimization of Hidden Objects:} To address the problem that hidden objects such as inside the box or inside vehicles usually have low contrast, enhance the visual recognition of hidden objects, and further help identify and analyze possible abnormal objects or contraband.

\textbf{(2). Enhancement of Spatial Edge Structures:} The goal is to improve the original stereoscopic  and edge clarity of objects, restoring their original structural features.

\textbf{(3). Crack Detection:} In some raw data, noise may obscure important crack details. This test aims to clearly restore these details for easier identification and analysis.

\textbf{(4). Test Piece Analysis:} Utilizing professional test pieces to evaluate the model's performance under specific technical indicators, such as resolution, demonstrating the effectiveness of the BSoNet algorithm.

Through these tests, we can accurately assess the optimization effects of BSoNet in practical applications, ensuring that it can effectively improve the applicability and reliability of PBI. Fig. \ref{fig4} shows six sets of optimization results for these representative cases, with the scanning parameters for tube voltage and tube current indicated.  Our method is highlighted in red boxes, and each group is sequentially labeled with the letters a-e. The relative positions of different algorithms are marked in group (a). Red arrow regions indicate key improvement areas, while yellow and green arrow regions highlight specific details. The lower right corner of each visual optimization result also details the corresponding local contrast values.

Groups (a) and (b) simulate scenarios where potential threat objects are hidden. Group (b) simulates common risk situations encountered in routine security inspections, where a variety of objects, including bottles, are placed inside a box. Due to differences in material, density, and shape, these objects exhibit varying contrast levels in backscatter imaging. In security inspection scenarios, container-type objects (such as bottles) are typically considered potential threats, as they may conceal liquid contraband, which could include flammable, explosive, or toxic substances. Therefore, accurately detecting such items is crucial. After optimization by BSoNet, the contrast of the bottle, highlighted in the red region, is significantly enhanced, with edges becoming clearer and smoother. Group (a) simulates a scenario where contraband is hidden inside a tire. We choose white sugar as a substitute for contraband because its physical properties, especially its density and scattering characteristics, are similar to some contraband, such as drugs. The optimized imaging results using BSoNet also demonstrate enhanced contour and detail for the hidden object within the tire. Especially in the red region, BSoNet significantly improves contrast, making the hidden object more discernible. These two experiments demonstrate the practical  application value of BSoNet in applications such as routine security screenings and border inspections, where the identification of hidden contraband is critical.  BSoNet helps rapidly and accurately identify potential security threats, thereby improving detection efficiency and accuracy.

In the test setup of group (c), the scenario is designed to demonstrate BSoNet's capability in handling images with a visually prominent foreground and segmented background. This environment places a water bottle at the corner of a table, forming a clear foreground, while the background is divided into two sections. This layout challenges the algorithm to precisely enhance the spatial structure of the foreground object while effectively distinguishing and optimizing the complex background behind it. Through BSoNet's processing, the red-marked area highlights the 3D structure of the water bottle, showcasing our algorithm's superior ability to enhance object edges and details. The yellow-marked area precisely restores the structure of the table corner. Meanwhile, the green-marked area enhances the contrast between the two backgrounds and clearly distinguishes the different backgrounds. These improvements not only enhance the visual clarity and layering of the image but also demonstrate BSoNet's effectiveness in enriching image information, showing its excellent capability in handling details and improving contrast in scenes.

In some industrial non-destructive testing scenarios, PBI can be used to inspect defects in large structures, back-wall containers, or multi-layer composite materials, particularly cracks and voids. Therefore, BSoNet needs to have the ability to effectively optimize image quality in such environments. Group (d) demonstrates BSoNet's  ability to handle crack regions that are difficult to identify due to occlusion, which is particularly crucial in actual applications such as internal structure integrity maintenance inspections. As shown in the red-marked areas, small cracks are obscured by noisy backgrounds, posing a potential risk to the structural safety or functional integrity. BSoNet effectively reduces noise interference and significantly improves the visibility of the crack region, demonstrating its ability to quickly and accurately identify structural defects in maintenance and inspection tasks that require precise image analysis. Deploying BSoNet in industrial sites is expected to reduce maintenance costs, provide high-quality images to support more precise safety assessments, and ultimately improve overall production and safety management levels.

In group (e), the polyethylene rod detection scenario is used to verify the recognition and optimization capabilities of different algorithms for specific targets. In the raw data, the polyethylene rod is not clearly visible due to noise interference, posing a challenge in enhancing the visibility of the target polyethylene rod (as shown in the red-marked areas) while effectively controlling background noise and reducing visual distortion. The results show that BSoNet significantly improves the clarity and recognizability of the polyethylene rod while maintaining its overall structure.  In comparison, other processing algorithms, although attempting to enhance target recognition, introduce other visual issues. For example, bilateral filtering makes the overall image more blurred, UNet causes significant morphological distortion, and Uformer fails to effectively optimize image noise. This test highlights BSoNet's excellent capabilities in feature enhancement and noise control.

In the test of group (f), we focus on evaluating and verifying BSoNet's ability to enhance the recognizability of polyethylene line pairs (as shown in the red-marked areas) against an air background, thereby testing its performance in maintaining spatial resolution in backscatter images. This is the key factor for BSoNet to accurately distinguish object features in practical applications. Through this test, BSoNet not only accurately restored the polyethylene line pairs but also significantly improved the overall image quality. Specifically, other parts of the polyethylene were processed more smoothly, and the boundaries between the polyethylene and the air background became sharper and clearer. This test demonstrates BSoNet's outstanding capability in maintaining image spatial resolution quality.

\begin{figure}[!ht]
\centerline{\includegraphics[width=\columnwidth]{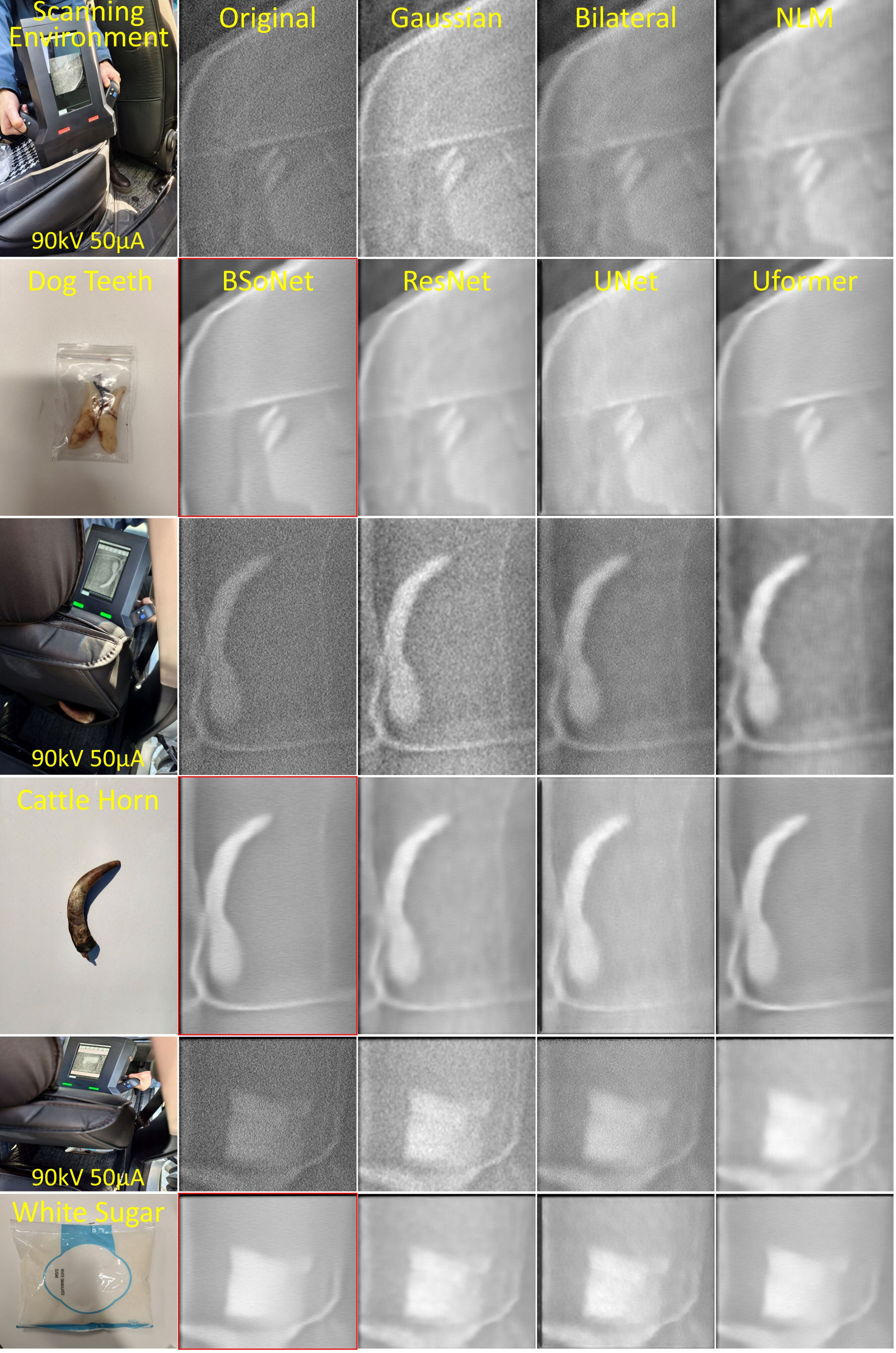}}
\caption{A simulated scenario of hiding contraband in car seats during border inspection, where dog teeth and cattle horns serve as substitutes for smuggled animal products, and white sugar is used to simulate  drugs.}
\label{fig7}
\end{figure}

Border inspections typically require the rapid scanning of unconventional areas within vehicles to detect  contraband (such as drugs, smuggled goods, etc.) that may be carried inside. PBI is widely used in such scenarios due to its flexibility. However, the complex layout of vehicle interiors increases the difficulty of detection. Therefore, BSoNet needs to enhance the visibility of targets in such environments, particularly in hiding spaces such as car seats, tires, and door panels, making potential contraband easier to identify.
Fig. \ref{fig7} illustrates a common scenario in border inspections, where contraband is concealed within a car seat. To further enhance the authenticity of the experiment, real-world photographs of the scanning environment were provided. In addition to the sugar used in group (a), cattle horn and dog teeth were selected as substitutes for smuggled animal products. Larger contraband substitutes (such as white sugar and cattle horns) were placed behind the front passenger seat and scanned from the front, while smaller substitutes (such as dog teeth), were inserted under the rear seat for scanning. We applied algorithmic optimization to the acquired backscatter images and compared the results before and after optimization, which also fully demonstrated the outstanding ability of BSoNet in enhancing image quality and improving detection performance. Deploying BSoNet in border inspection systems is expected to shorten the inspection process, reduce labor costs and false negative rates, while providing clearer image-based evidence for subsequent risk assessments.

Due to the inability to obtain ideal noise-free reference images, the quantitative analysis must rely on the backscatter images themselves. Moreover, backscatter images differ from natural images in terms of dynamic range, noise distribution, and other characteristics, making it necessary to select appropriate quantitative metrics for evaluation. We have introduced CPBD (Cumulative Probability of Blur Detection) as a quantitative evaluation metric for backscatter images~\cite{narvekar2011no}. CPBD, based on the characteristics of human blur perception, can effectively measure image sharpness. This metric detects the probability of blur at edges in the image and calculates the cumulative probability to quantitatively assess the degree of blur. A higher CPBD value indicates a sharper image with less blur, while a lower value indicates a blurrier image with poor sharpness. Since blur in backscatter images directly affects the recognizability of content and subsequently impacts follow-up tasks, CPBD can provide precise evaluation of image sharpness and blur, thereby effectively supporting the optimization and analysis of backscatter image quality.

In addition, we have defined a local contrast (${{C}_{l}}$) measurement  method that evaluates image contrast based on the intensity differences between each pixel and its eight neighboring pixels. Specifically, we calculate the sum of the squared  intensity differences between each pixel and its surrounding pixels, then accumulate and average these differences, followed by taking the square root, to estimate the overall contrast of the image to some extent. We define local contrast as~\eqref{eq2}:
\begin{equation}
\resizebox{0.9\columnwidth}{!}{$
{{C}_{l}}=\sqrt{\frac{\sum\limits_{i=1}^{M}{\sum\limits_{j=1}^{N}{\left( \sum\limits_{(k,l)\in N(i,j)}{(f(}i,j)-f(k,l){{)}^{2}} \right)}}}{8\times (M-2)\times (N-2)+5\times (2\times (M-2)+2\times (N-2))+3\times 4}},
$}
\label{eq2}
\end{equation}
Where, $M$ and $N$ represent the number of rows and columns of the image, respectively. $f(i,j)$ denotes the pixel value of the image at position $(i,j)$. $(k,l)$ are the coordinates of the neighboring pixels around position $(i,j)$, and $f(k,l)$ represents the pixel value at position $(k,l)$. 

\begin{figure}[!t]
\centerline{\includegraphics[width=\columnwidth]{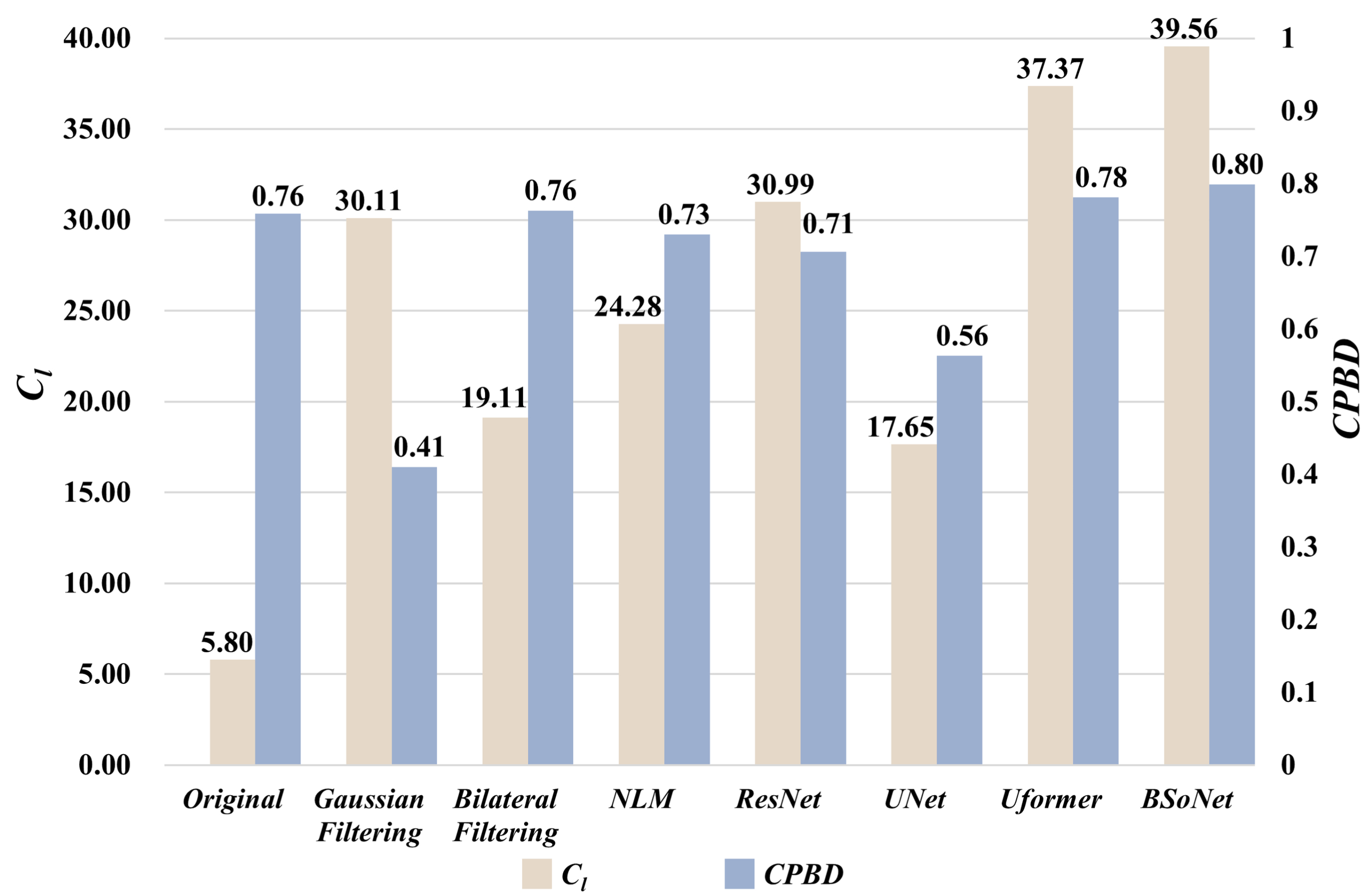}}
\caption{The average quantitative results of ${{C}_{l}}$ and CPBD in the test dataset.}
\label{fig8}
\end{figure}

Fig. \ref{fig8} shows the average quantitative results of CPBD and ${{C}_{l}}$ for all methods in the test dataset. Combined with the visual optimization effects shown in Fig. \ref{fig4} and Fig. \ref{fig7}, our BSoNet  demonstrates consistent performance across different scenarios, the optimized image quality has obvious competitive advantages. Each set clearly proving BSoNet's high efficiency in enhancing image contrast, refining object edges, reducing noise, and improving overall image clarity and resolution. Whether in improving the accuracy of security inspections or in the application of industrial safety inspections, BSoNet has shown excellent performance, effectively meeting the demands for high-quality image processing in various fields. These results strengthen the technical foundation of our PBI and lay the groundwork for further application and development of BSoNet in the field of image processing.

To further validate the feasibility of BSoNet in practical applications, we evaluated both the training and inference efficiency of the model.  Under current experimental environment, the average training time of BSoNet is 492.62 seconds per epoch. In the remote computing scheme (detailed in Chapter VI ``MODEL DEPLOYMENT"), the average time from the server receiving client data to completing inference and returning the result is only 0.43 seconds. Additionally, BSoNet comprises approximately 30.24 million trainable parameters. These results indicate that BSoNet can achieve high-quality image optimization while meeting the real-time requirements of PBI.

\subsection{Robustness Evaluation}
\begin{figure}[!ht]
\centerline{\includegraphics[width=\columnwidth]{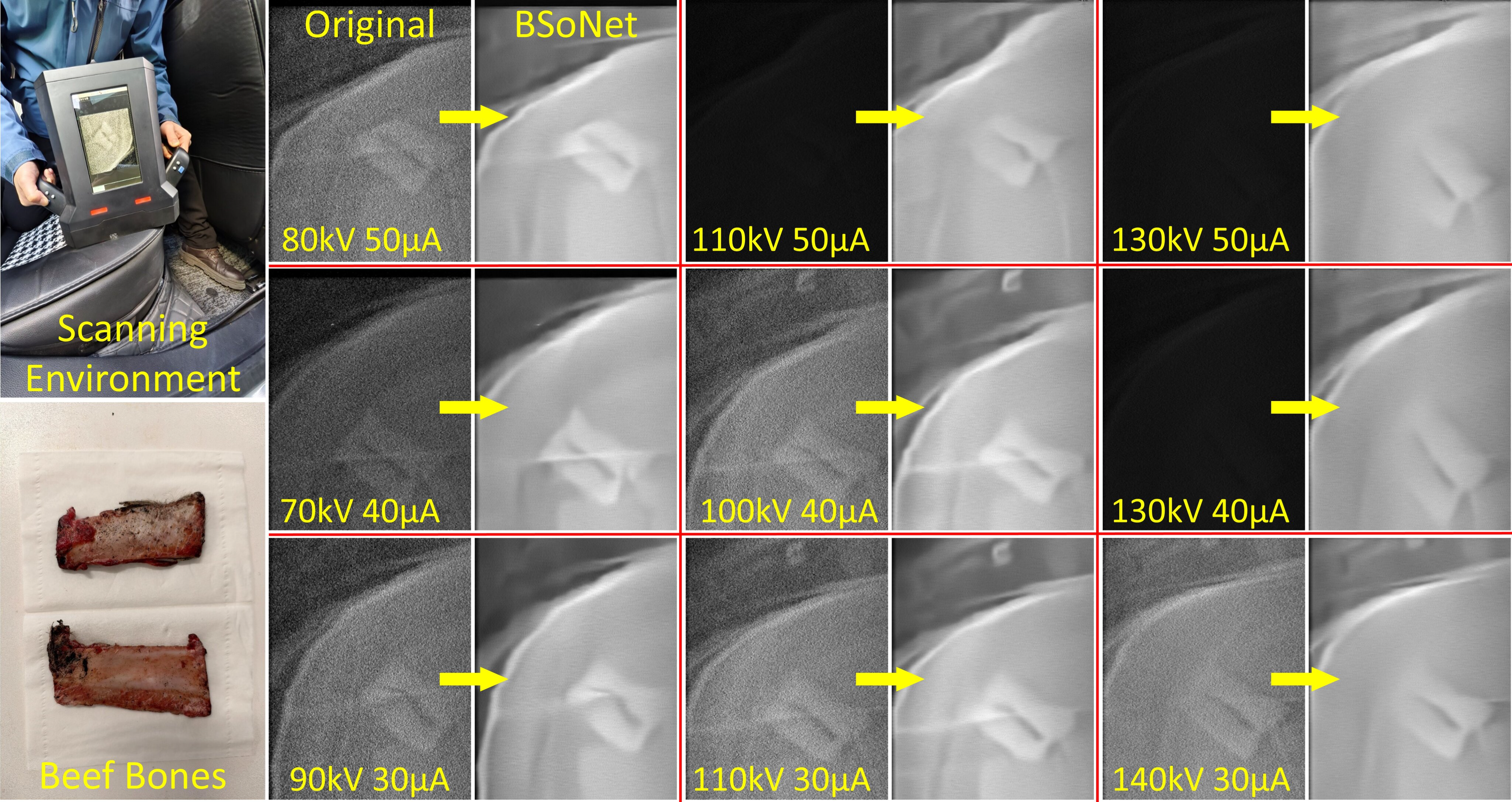}}
\caption{Robustness validation experiment of BSoNet. It shows the original images captured under different scanning conditions and their corresponding results after optimization by BSoNet.}
\label{fig6}
\end{figure}

We designed a set of experiments to evaluate the robustness of BSoNet under different imaging conditions. We placed cow bone inside the car seat, as shown in Fig. \ref{fig6}, and conducted scans under various imaging conditions, such as tube voltage, tube current, and others. Since it is impossible to ensure that the scanning speed, duration, and device status are exactly the same in each scan during actual operation, the randomness of these three parameters is introduced into the robustness evaluation experiments, resulting in different imaging resolutions for each scan. To ensure consistency in the presentation of results, all images are adjusted to the same size for display.

We simulated these uncertainty factors and displayed nine sets of experimental results for examples. Each set includes the original image, the corresponding imaging parameters, and the optimized image after BSoNet processing, as shown in Fig. \ref{fig6}. These results demonstrate that BSoNet provides nearly consistent optimization performance under different scanning conditions, validating its outstanding robustness.

\section{Model Deployment}
This section provides guidance on the deployment of BSoNet on PBI. Considering that PBI's hardware resources are generally limited and computing power is insufficient, deep learning models usually require strong computing power and large memory. Even when lightweight models can be deployed on local devices in some cases, the inference time often fails to meet the real-time requirements of PBI, and the performance can not meet the optimization quality requirements. In this context, we recommend deploying BSoNet through remote communication. A communication schematic in a common situation is shown in Fig. \ref{fig5}. The deployment through remote communication offers the following advantages:

\begin{figure}[!ht]
\centerline{\includegraphics[width=\columnwidth]{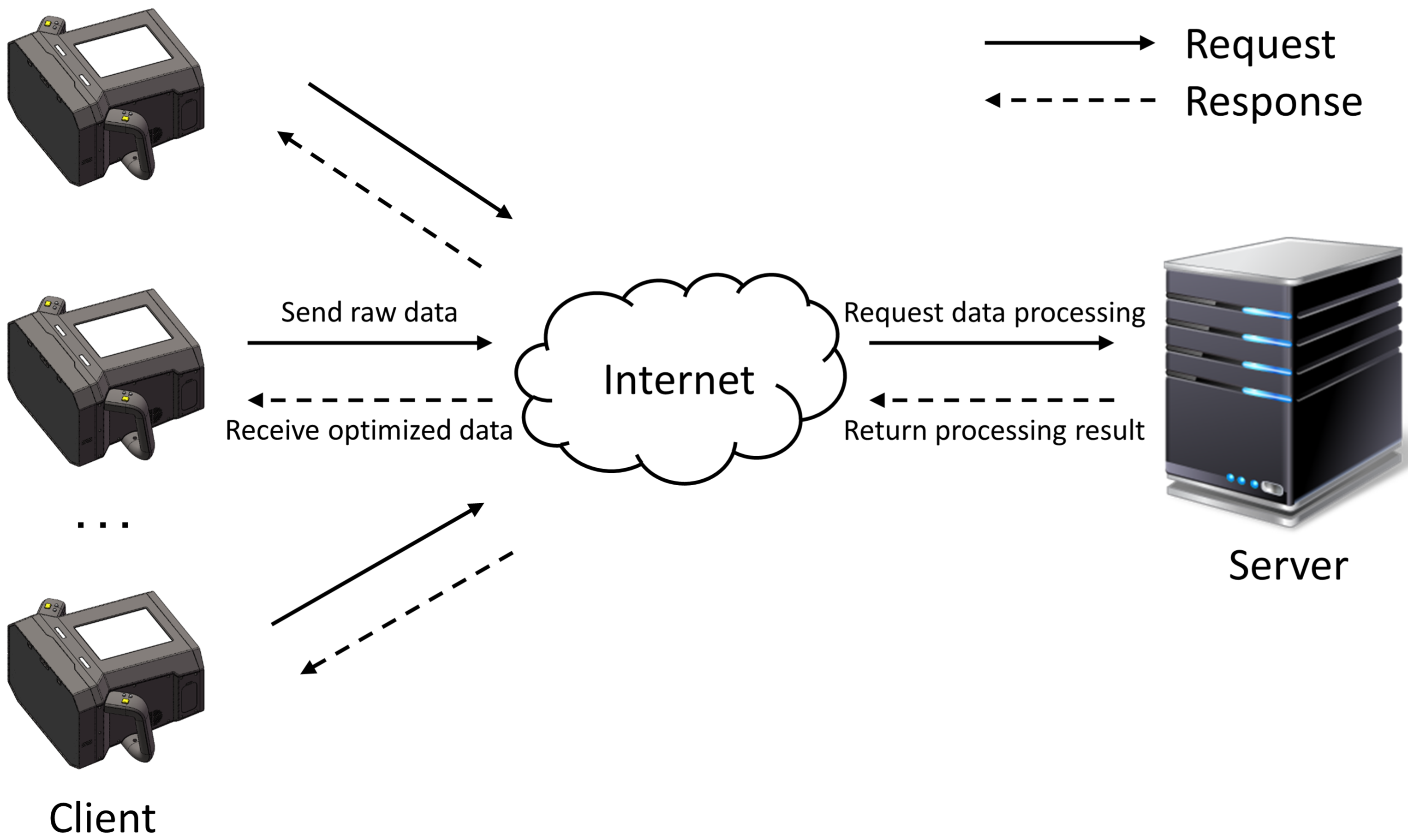}}
\caption{A schematic diagram of remote communication for PBI. The client sends raw data and receives optimized data, while the server processes the data and returns the results.}
\label{fig5}
\end{figure}

\textbf{1. Performance Improvement:} Servers typically have more powerful computing capabilities, which improves the operation efficiency of the model.

\textbf{2. Resource Conservation:} Avoid high computational processing on PBI, save power consumption and extend the life of the device.

\textbf{3. Ease of Maintenance:} Centralized management and updating of the model on the server facilitate maintenance and upgrades.

\textbf{4. Cost Reduction:} Lower hardware requirements and costs for the PBI, improves the market competitiveness of products, and increases the company's profit margin.

Our system architecture is mainly composed of two parts, the server and the client, the respective functions are summarized as follows:

\textbf{Server Architecture:} The server primarily includes gRPC services~\cite{grpc2018high}, model loading and prediction, and data storage and management modules. The gRPC services are responsible for receiving and processing data requests from the client; the model loading and prediction module is used to load the pre-trained BSoNet model and process the data transmitted from the client, optimizing backscatter images; the data storage and management module is responsible for storing the raw data and the processed image results on the server, managing them through appropriate paths and naming conventions.

\textbf{Client Architecture:} The client is mainly responsible for data collection, transmission, and receiving and displaying the optimization results.  The client packages the acquired raw image data and transmits it to the server via gRPC.  After the server completes the processing, the optimized backscatter image is then 
 returned to the client via gRPC, where it is displayed.

Deploying through remote communication allows PBI to fully utilize the computing power of the server, making the management and maintenance of device algorithms more convenient while also reducing overall costs.

\section{Conclusion}
This paper presents BSoNet, a novel deep learning framework designed specifically for optimizing the image quality of PBI. In the implementation of BSoNet, the Resolution Adaptive Network (RANet) ensures consistency in image dimensions and structural information before and after optimization through end-to-end adaptive adjustments. This allows the model to maintain the original image size under different scanning conditions while providing consistent optimization quality, effectively accommodating the variable environments of practical applications. BSformer comprehends image content on a global scale and finely optimizes details and contrast on a local level, significantly enhancing target features within the image and overall optimizing noise.

We also introduce an enhanced self-supervised learning strategy, with additional noise augmentation on the  raw backscatter data during the training process. This compels BSoNet to actively adapt to and optimize complex noise conditions, significantly improving the model's adaptability and processing ability to the inherently high noise and low signal quality environment in the backscatter image. Consequently, this strategy enhances the performance of PBI in real-world usage scenarios.

The experimental results demonstrate that BSoNet effectively handles various complex noises and artifacts encountered in practical backscatter imaging, thereby generating clearer and more recognizable high-quality images. This enables BSoNet to show great potential in areas such as public safety and industrial inspection. Therefore, BSoNet marks a significant advancement in portable backscatter imaging technology, driving the development of PBI.

However, there are also some limitations in this study. First, Noise2Void has a strong dependence on specific tasks during training. Specifically, during training, we applied noise pre-augmentation (standard Gaussian noise with an intensity of 400) to the training data generated by Noise2Void, aiming to help the model learn specific noise patterns and enhance its denoising capability. However, this noise pre-augmentation strategy may lead to performance bottlenecks and lack sufficient adaptability when the model faces other types of noise. Additionally, as the core mechanism of the Noise2Void prior task, random masking and replacement relies on the assumption that neighboring pixels in the image have strong correlations. However, in scenarios with higher noise levels or weaker signals, this assumption may not hold, which can impair the model's generalization ability in these complex situations.

Based on the above studies, future work will focus on optimizing the self-supervised training strategy to enhance the model's generalization ability and adaptability in this field. A key area of focus will be improving noise modeling techniques, particularly addressing the issue of strong dependency on the specific noise type caused by the noise pre-augmentation in the current strategy, and exploring more universal noise modeling approaches to avoid excessive reliance on a single noise pattern. Furthermore, to address the issue of potentially weak correlation between adjacent pixels, future work will also focus on designing more flexible and diverse training strategies to further improve performance in real-world applications. We will continue to deepen and optimize related research to provide more reliable and efficient solutions for this field and drive the further development of backscatter imaging technology.

\section{ACKNOWLEDGMENTS}
This work was supported in part by the Natural Science Foundation
of Zhejiang Province, [Grant/Award Number: LZ23A050002]; National
Natural Science Foundation of China, [Grant/Award Number: 12175012].

\bibliographystyle{IEEEtran}
\bibliography{refs}

\end{document}